\documentclass[twocolumn,showpacs,preprintnumbers,amsmath,amssymb,pre,aps]{revtex4}
\usepackage{graphicx}% Include figure files
\usepackage{dcolumn}% Align table columns on decimal point
\usepackage{bm}% bold math

\begin{document}
\title{Imaginary chemical potential quantum Monte Carlo for Hubbard molecules}
\author{Fei Lin, Jurij \v{S}makov, Erik S. S\o{}rensen, Catherine Kallin and A. John Berlinsky}
 \affiliation{Department of Physics and Astronomy, McMaster University, Hamilton, Ontario,
 Canada L8S 4M1}
 \date{\today}

\begin{abstract}
We generalize the imaginary chemical potential quantum Monte Carlo
(QMC) method proposed by Dagotto \emph{et al.} [Phys. Rev. B {\bf
41}, R811 (1990)] to systems without particle-hole symmetry. The
generalized method is tested by comparing the results of the QMC
simulations and exact diagonalization on small Hubbard molecules,
such as tetrahedron and truncated tetrahedron. Results of the
application of the method to the C$_{60}$ Hubbard molecule are
discussed.
\end{abstract}

\maketitle
\section{Introduction}
Knowledge of the evolution of energy levels with doping in strongly
correlated systems is of significant importance for understanding the
physical mechanisms leading to their unconventional properties. For
example, information about the changes in the ground state energy of
model electron systems (such as Hubbard model) upon electron or hole
doping may be used to confirm or disprove hypotheses about the origin
of the pairing mechanism, eventually leading to superconductivity. As
accessing this information analytically usually requires the use of
various approximations, numerical techniques are often the only tools
which can provide unbiased estimates for the observables of interest.

We have recently applied the auxiliary field quantum Monte Carlo
(AFQMC) on a C$_{60}$ molecule to extract the electronic binding
energies \cite{lin04}. AFQMC has been widely used in Hubbard
Hamiltonian simulations since its introduction by Blankenbecler
\emph{et al.} \cite{sugar81a,sugar81b}, and its further
development by Hirsch \cite{hirsch83a83b85} and White \emph{et
al.} \cite{white89}. Being a finite-temperature technique, AFQMC
does not allow easy access to the physical observables, not
represented by thermodynamic averages, such as energy gaps. A
convenient procedure to extract this additional information from
the AFQMC data was proposed by Dagotto \emph{et al.}
\cite{dagotto90}, who introduced imaginary chemical potentials in
AFQMC simulations. It was then used to extract the charge gaps of
the one-band Hubbard model on finite two-dimensional (2D) square
lattices.

In the present paper we generalize this formalism to systems
without particle-hole symmetry, such as the tetrahedron, truncated
tetrahedron and C$_{60}$ molecules. The canonical partition
function ratios are obtained from the expansions of the AFQMC
determinant ratios for a set of finite temperatures $T$, which are
subsequently used to extract charge gaps at low temperatures. This
generalization results in the appearance of an extra phase factor
in the expansion of the determinant ratios, which reduces to unity
in the particle-hole symmetric systems.

The rest of the paper is organized as follows. First, we briefly
describe the imaginary chemical potential QMC (ICPQMC) formalism. Then
simulation results on some Hubbard molecules are presented to
illustrate our method. The results are compared with the data obtained
by exact diagonalization (ED) on small molecules and projector QMC
(PQMC) on larger ones.

\section{Methodology}
We start with an expansion of the grand canonical partition
function $Z_{\text{GC}}(\mu)$ in terms of canonical partition
functions $Z_{\text{C}}(n)$ \cite{dagotto90}:
\begin{equation}
Z_{\text{GC}}(\mu)=\text{Tr}\,e^{-\beta(\hat{H}-\mu\hat{N})}=e^{\beta\mu
N}\sum_{n=-N}^{N}e^{\beta\mu n}Z_{\text{C}}(n), \label{gcexpand}
\end{equation}
where $\beta=1/(k_BT)$ is the inverse temperature, $\mu$ is the
chemical potential, $n$ is the deviation of the particle number from
half-filling (positive or negative, denoting electron or hole doping,
respectively) in canonical ensemble, $\hat{N}$ is the electron number
operator, and $N$ is the number of spatial lattice sites in the
system.  $\hat{H}$ is the usual one-band Hubbard Hamiltonian:
\begin{equation}
H=-\sum_{\langle
ij\rangle\sigma}t_{ij}(c_{i\sigma}^{\dagger}c_{j\sigma}+h.c.)+U\sum_{i}n_{i\uparrow}
n_{i\downarrow}-\frac{U}{2}\sum_{i\sigma}n_{i\sigma}.
\label{hubbard}
\end{equation}
The summation in the hopping term $t_{ij}$ is performed over all
nearest-neighbor pairs of the Hubbard molecule. For the C$_{60}$
molecule we have set $t_{ij}=t$ for the links between the pentagons
and hexagons and $t_{ij}=1.2t$ for the links between hexagons. In all
other cases $t_{ij}$ was set equal to $t$ for all links, with $t$ used
as an energy unit. $U$ is the on-site Coulomb repulsion (Hubbard)
term, and an extra diagonal term has been added to the Hamiltonian so
that $\mu=0$ corresponds to half-filling on bipartite lattices.

Following Dagotto \emph{et al.} \cite{dagotto90}, we analytically
continue Eq.\ (\ref{gcexpand}) to imaginary chemical potential
$\mu\rightarrow i\lambda$, where $\lambda$ is real. Then the
inverse Fourier transform of Eq.\ (\ref{gcexpand}) gives
\begin{equation}
Z_{\text{C}}(n)=\frac{\beta}{2\pi}\int_0^{2\pi/\beta}d\lambda\,
e^{-i\beta\lambda(n+N)}Z_{\text{GC}}(\mu=i\lambda).
\label{zinverse}
\end{equation}
In AFQMC the grand canonical partition function is given by
\begin{eqnarray}
Z_{\text{GC}}(\mu)&=&\sum_{\{\sigma\}}\prod_{\alpha=\pm 1}\det
[1+B_{L}(\alpha)B_{L-1}(\alpha)\cdots
B_{1}(\alpha)]\nonumber\\
&=&\sum_{\{\sigma\}}\det O(\{\sigma\},\mu)_{\uparrow}\det
O(\{\sigma\},\mu)_{\downarrow}, \label{zfinal}
\end{eqnarray}
where the fermion degrees of freedom have been traced out, and the
$B_l$ matrices are defined as
\begin{eqnarray}
B_{l}(\alpha)&=&e^{-\Delta\tau K}e^{V^{\alpha}(l)},\label{bdefine}\\
(K)_{ij}&=&\left\{\begin{array}{cc}
-t_{ij} & \text{for $i$,$j$ nearest neighbors},\\
0 & \text{otherwise},
\end{array}\right.\\
V_{ij}^{\alpha}(l)&=&\delta_{ij}[\gamma\alpha\sigma_{i}(l)+\mu\Delta\tau].
\label{vls}
\end{eqnarray}
Here $\Delta\tau$ is the imaginary time discretization interval
and $\tanh^2(\gamma/2)=\tanh(\Delta\tau U/4)$. Thus, the original
problem of taking a trace over fermionic degrees of freedom has
been replaced by a problem of tracing over auxiliary Ising
variables $\sigma_i(l)$, introduced at every space-time point $(i,
l\Delta\tau)$. Inserting Eq.\ (\ref{zfinal}) into Eq.\
(\ref{zinverse}), we get
\begin{eqnarray}
Z_{\text{C}}(n)&=&\sum_{\{\sigma\}}\frac{\beta}{2\pi}\int_0^{2\pi/\beta}d\lambda\,
e^{-i\beta\lambda(n+N)}\det O(\{\sigma\},i\lambda)_{\uparrow}\nonumber\\
& &\times \det
O(\{\sigma\},i\lambda)_{\downarrow}.\label{zinverse1}
\end{eqnarray}
\begin{figure}
  \centering
  % Requires \usepackage{graphicx}
  \begin{tabular}{c}
  \resizebox{70mm}{!}{\includegraphics{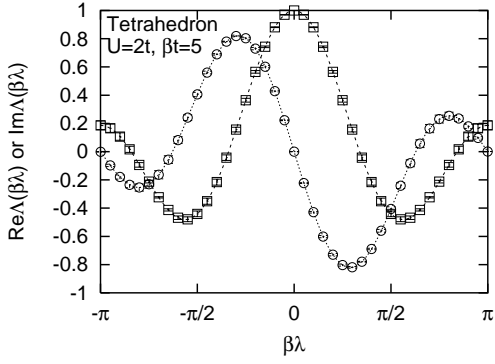}} \\
  \end{tabular}
  \caption{Fit of real (squares) and imaginary (circles) parts of
  determinant ratios according to Eq.\ (\ref{detexpand}) for a
  tetrahedron molecule.}
  \label{c4u2b5det}
\end{figure}
\begin{figure}
  % Requires \usepackage{graphicx}
  \begin{tabular}{c}
    % after \\: \hline or \cline{col1-col2} \cline{col3-col4} ...
    \resizebox{70mm}{!}{\includegraphics{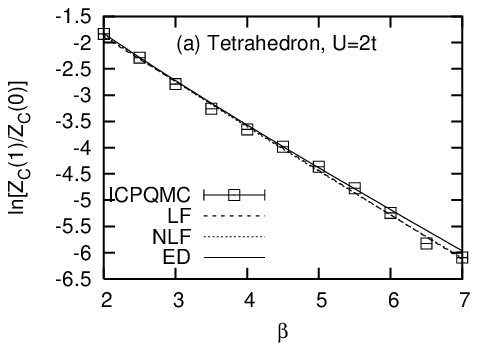}} \\
    \resizebox{70mm}{!}{\includegraphics{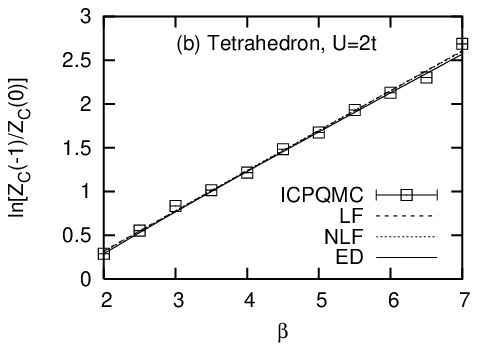}} \\
  \end{tabular}
  \caption{Fits of ICPQMC data for a tetrahedron molecule at low
  temperatures for electron (a) and hole (b) doping. NLF forms for (a)
  and (b) are given by Eq.\ (\ref{nlf}) and Eq.\ (\ref{lne10swfit}). ED
  results are also shown for comparison.}\label{c4u2lnl}
\end{figure}
Dividing Eq.\ (\ref{zinverse1}) by $Z_{\text{GC}}(\mu=0)$ from
Eq.\ (\ref{zfinal}) yields
\begin{equation}
\frac{Z_{\text{C}}(n)}{Z_{\text{GC}}(0)}=\sum_{\{\sigma\}}P(\{\sigma\},0)\frac{\beta}{2\pi}
\int_0^{2\pi/\beta}d\lambda e^{-i\beta\lambda n}\Lambda(\beta
\lambda)\label{znf}
\end{equation}
where
\begin{equation}\label{biglambda}
\Lambda(\beta\lambda)\equiv e^{-i\beta\lambda N}\frac{\det
O(\{\sigma\},i\lambda)_{\uparrow}\det
O(\{\sigma\},i\lambda)_{\downarrow}}{\det
O(\{\sigma\},0)_{\uparrow}\det O(\{\sigma\},0)_{\downarrow}},
\end{equation}
and
\begin{equation}
P(\{\sigma\},0)=\frac{\det O(\{\sigma\},0)_{\uparrow}\det
O(\{\sigma\},0)_{\downarrow}}{\sum_{\{\sigma\}}\det
O(\{\sigma\},0)_{\uparrow}\det O(\{\sigma\},0)_{\downarrow}}
\end{equation}
is the probability distribution for $Z_{\text{GC}}(0)$. Since
there is always an energy gap above and below half-filling for any
finite system, we expect the fermion determinants to be nearly
$\lambda$-independent at low temperatures. Therefore, we can
generate the Ising field configurations $\{\sigma\}$ for
$\lambda=0$ and use these configurations to calculate system
properties at $\lambda\neq 0$.

Similar to the expansion in Eq.\ (\ref{gcexpand}), we expect that
the determinant ratio $\Lambda(\beta\lambda)$ in Eq.\
(\ref{biglambda}) can be expressed as complex Fourier series in
the particle number $n$
%\begin{widetext}
\begin{eqnarray}
\Lambda(\beta\lambda)&=&c_0(\{\sigma\})+\sum_{n=1}^N[c_n(\{\sigma\})+c_{-n}(\{\sigma\})]\cos(\beta\lambda
n)\nonumber\\
&&+i\sum_{n=1}^N[c_n(\{\sigma\})-c_{-n}(\{\sigma\})]\sin(\beta\lambda
n), \label{detexpand}
\end{eqnarray}
%\end{widetext}
where $n$ ($-n$) represents a doping of $n$ electrons (holes) with
respect to half-filling. Eq.\ (\ref{detexpand}) is real for
systems with particle-hole symmetry, since then
$c_n(\{\sigma\})=c_{-n}(\{\sigma\})$. When we substitute Eq.\
(\ref{detexpand}) back into Eq.\ (\ref{znf}), we see that
$Z_{\text{C}}(n)/Z_{\text{GC}}(0)=\langle c_{n}\rangle$, where the
average $\langle\ldots\rangle$ is over the Ising field
configurations $\{\sigma\}$ generated from the probability
distribution $P(\{\sigma\},0)$. In the case of negative weight we
replace $P$ by its absolute value $|P|$ and include a sign
$S=P/|P|$ in the average:
$Z_{\text{C}}(n)/Z_{\text{GC}}(0)=\langle c_{n}S\rangle/\langle
S\rangle$. The average $\langle\ldots\rangle$ now refers to the
probability distribution $|P|$. As we are interested in regions
near half-filling, the sign problem does not limit the
applicability of our method.

At low temperatures the canonical partition function ratio will be
dominated by $\Delta_{n,0}=E(n)-E(0)$, the energy difference between
the ground states for the two fillings, and will take the form
\cite{dagotto90}
\begin{equation}
\frac{Z_{\text{C}}(n)}{Z_{\text{C}}(0)}=\frac{\langle
c_n\rangle}{\langle c_0\rangle}\rightarrow
d_{n,0}e^{-\beta\Delta_{n,0}}\quad\text{as}\quad
\beta\rightarrow\infty,\label{lf}
\end{equation}
where $d_{n,0}=d_n/d_0$, with $d_n$ being the degeneracy of the
ground state at filling $n$. When there is an energy level close
to the ground state (produced by the elementary excitations such
as spin waves), we include it explicitly in the fitting
expressions:
\begin{eqnarray}
\frac{Z_{\text{C}}(n)}{Z_{\text{C}}(0)}=\frac{\langle
c_n\rangle}{\langle c_0\rangle}&\rightarrow&
\frac{d_{n,0}e^{-\beta\Delta_{n,0}}+f_{n,0}e^{-\beta\Delta_{\text{sw}}^n}}{1+f_{0,0}
e^{-\beta\Delta^{0}_{\text{sw}}}}\nonumber\\ &\text{as}&\quad
\beta\gg\frac{1}{\Delta_{n,0}},\frac{1}{\Delta^{n}_{\text{sw}}}.\label{nlf}
\end{eqnarray}
Here $f_{n,0}=d^n_{\text{sw}}/d_0$, where $d^n_{\text{sw}}$ is the
degeneracy of the spin wave state at filling $n$, and
$\Delta^n_{\text{sw}}=E_{\text{sw}}(n)-E(0)$ are the spin-wave
gaps at filling $n$ with respect to the ground state at
half-filling.

\begin{table}
  \centering
  \begin{ruledtabular}
  \begin{tabular}{|l|l|ccc|}
   & & LF & NLF & ED (PQMC) \\
  \hline
  C$_4$           & $\Delta_{1,0}$ & 0.85(1)     & 0.87(2) & 0.82843  \\
  $(U=2t)$        & $\Delta_{-1,0}$ & -0.458(9)  & -0.358(9)  & -0.34949 \\
                  & $\Delta_{2,1}$ & 1.9(1)      & 2.0(1) & 2.0 \\
                  & $\Delta_{-2,-1}$ & -0.400(6) & -0.397(6) & -0.40466 \\
  \hline
  C$_8$           & $\Delta_{\pm 1,0}$ & 1.20(2)  & 1.27(2) &  1.26224 \\
  $(U=4t)$        & $\Delta_{\pm 2,\pm 1}$ & 1.44(3) & 1.5(2) & 1.27490  \\
  \hline
  C$_{12}$        & $\Delta_{1,0}$   & 0.81(1)  & 0.997(7) & 0.99596 \\
  $(U=2t)$        & $\Delta_{-1,0}$  & 0.041(1) & 0.10(1)  & 0.07408 \\
  \hline
  C$_{60}$        &$\Delta_{1,0}$ & 0.43(5) & 0.57(3)  &  $0.561(7)^{*}$ \\
  $(U=4t)$        &$\Delta_{-1,0}$ & 0.9(1) & 0.88(4)  & $0.86(2)^{*}$ \\
  \end{tabular}
  \end{ruledtabular}
   \caption{ICPQMC results on tetrahedron (C$_4$), cube (C$_8$),
  truncated tetrahedron (C$_{12}$) and C$_{60}$
  molecules. $\Delta_{n,0}=E(n)-E(0)$ is the energy difference between
  the ground states of the two fillings. Data marked by $*$ are
  PQMC results, as described in Section
  \ref{sectionC60}.}\label{molecules}
\end{table}
Based on the above discussion, we formulate the following
calculation procedure:
\begin{enumerate}
    \item Generate the Ising field configuration $\{\sigma\}$ according to
    the probability distribution $P(\{\sigma\},0)$ in the AFQMC simulation
    of $Z_{\text{GC}}(0)$.
    \item Evaluate the average of the determinant ratio on the
    left-hand side of Eq.\ (\ref{detexpand}) over the Ising field
    configurations for a set of $\lambda$ values.
    \item Fit the real and imaginary parts of Eq.\ (\ref{detexpand})
    respectively to determine the average values of $\langle
    c_{n}\rangle$, $n=0,\pm 1,\pm 2, \ldots$. From now on we will
    refer to these values simply as $c_n$.
    \item Fit the canonical partition function ratios
    $Z_{\text{C}}(n)/Z_{\text{C}}(0)=c_n/c_0$,
    $n=\pm 1,\pm 2, \ldots$ with low temperature canonical partition
    function ratio expressions (\ref{lf}) or (\ref{nlf}) to obtain
    energy gaps. Below we will refer to these fits as linear (LF) and
    non-linear (NLF), respectively.
\end{enumerate}

\section{Application}
ICPQMC simulations have been carried out on tetrahedron, cube,
truncated tetrahedron and C$_{60}$ molecules. For each molecule,
we have run the simulations at temperatures $\beta t=2.0, 2.5,
\cdots, 6.5, 7.0$. The imaginary chemical potential $i\lambda$ was
chosen so that $0<\lambda<\pi/\beta$, and we used a set of 20
evenly distributed $\lambda$ values in this range. We have
numerically checked that the real part of Eq.\ (\ref{detexpand})
is an even function around $\lambda=0$, while the imaginary part
is odd. Using this property, we have mapped out the determinant
ratio data for $\lambda>0$. For the special particle-hole
symmetric case, such as a 2D square lattice, we have tested our
programs for the $2\times 2$ and $4\times 4$ systems, reproducing
the results of Ref.\ \cite{dagotto90}.

\begin{figure}
  \begin{tabular}{c}
    \resizebox{70mm}{!}{\includegraphics{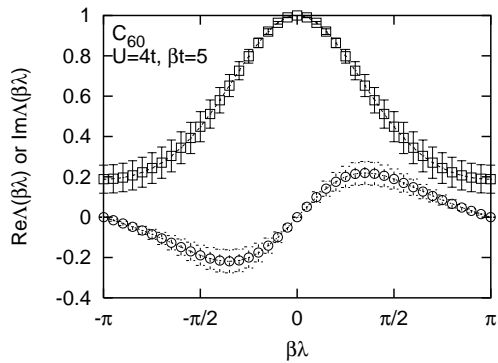}} \\
  \end{tabular}
  \caption{Fits of the real (squares) and imaginary (circles) parts of
  determinant ratios according to Eq.\ (\ref{detexpand}) for a
  C$_{60}$ molecule.}
  \label{c60b5det}
\end{figure}
\subsection{Tetrahedron, Cube and Truncated Tetrahedron}
Fig.\ \ref{c4u2b5det} displays the fit of the real and imaginary
parts of Eq.\ (\ref{detexpand}) for a tetrahedron molecule
(C$_4$). Similar fits were performed for data at all temperatures,
generating a set of coefficients $c_n$. Knowledge of these
coefficients enables us to calculate the partition function ratios
at various temperatures, and, at low temperatures, to linearly fit
the logarithm of these ratios to obtain the energy gaps and
degeneracy ratios. Results of this procedure are presented in
Fig.\ \ref{c4u2lnl} and Table \ref{molecules}. We see that
$\Delta_{2,1}$ and $\Delta_{-2,-1}$ from LF agree nicely with ED,
while $\Delta_{-1,0}$ does not. A likely cause of this discrepancy
is the existence of a highly degenerate ($d^0_{\text{sw}}=9$) spin
wave energy level very close ($\Delta^0_{\text{sw}}=0.14258t$) to
the ground state ($d_0=2$) at half-filling. Results of the fits
for electron ($n=1$) and hole ($n=-1$) doping are presented in
Figs.\ \ref{c4u2lnl}a and \ref{c4u2lnl}b respectively. The NLFs
were obtained using form (\ref{nlf}) by fixing the spin wave and
degeneracy parameters to the values found by ED:
\begin{equation}
\begin{array}{lcl}
f_{1,0}=3, & \quad &\Delta^1_{\text{sw}}=1.18268t, \\
f_{0,0}=4.5, & \quad & \Delta^{0}_{\text{sw}}=0.14258t.
\end{array}
%\ln[Z_{\text{C}}(1)/Z_{\text{C}}(0)]&=&\ln(d_{1,0}e^{-\beta\Delta_{1,0}}+3e^{-1.182676\beta})\nonumber\\
%& &-\ln(1+4.5e^{-0.14258\beta})
\label{lne10swfit}
\end{equation}
In the case of hole doping the spin wave term proportional to
$f_{-1,0}$ in the numerator of Eq.\ (\ref{nlf}) has been
neglected.

Examining Fig.\ \ref{c4u2lnl} we find that inclusion of spin waves
does not make a substantial difference: LF and NLF curves are
nearly overlapping [with a slight difference around $\beta t=2$
for $\ln(Z_{\text{C}}(1)/Z_{\text{C}}(0))$] and are very close to
the exact values. This insensitivity to the fitting parameters
makes accurate extraction of energy gaps from partition function
ratios difficult.

Analogous simulations and LF/NLF fittings were performed for cube
(C$_8$) and truncated tetrahedron (C$_{12}$) molecules. The energy
gaps measured using this procedure are summarized in Table
\ref{molecules}. The fitting procedure for C$_{60}$ is described
below.

\subsection{C$_{60}$}\label{sectionC60}
\begin{figure}
  \begin{tabular}{c}
  \resizebox{70mm}{!}{\includegraphics{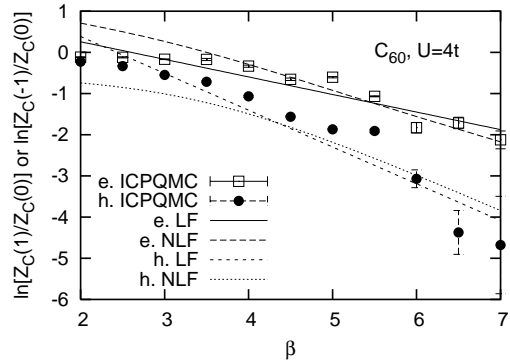}}\\
  \end{tabular}
  \caption{Fits of ICPQMC data for a C$_{60}$ molecule at low
  temperatures for electron and hole doping. e. (h.) denotes electron
  (hole) doping.}\label{c60fitgap}
\end{figure}

Fig.\ \ref{c60b5det} shows a fit of the real and imaginary parts
of Eq.\ (\ref{detexpand}) for a C$_{60}$ molecule. The imaginary
parts of the determinant ratios are positive for $\lambda>0$,
which is different from the tetrahedron and truncated tetrahedron
cases. The cause of this difference is the relative size of
electron and hole gaps in the system. The larger the gaps, the
smaller the corresponding canonical partition function ratios
$Z_{\text{C}}(n)/Z_{\text{C}}(0)$. For the tetrahedron and the
truncated tetrahedron the electron gaps are larger than the hole
gaps, so $c_1<c_{-1}$. In contrast, for the C$_{60}$ molecule the
electron gap is smaller than the hole gap, so $c_1>c_{-1}$.
Therefore, the relative size of the canonical partition function
ratios results in a positive imaginary part of the determinant
ratios for $\lambda>0$ in Fig.\ \ref{c60b5det} due to Eq.\
(\ref{detexpand}). Similar fits were done for other temperatures,
and the resulting canonical partition function parameters $c_n$
were obtained to calculate the canonical partition function
ratios. Unfortunately, Eq.\ (\ref{nlf}) contains too many fitting
parameters to provide unique fits to the data. Therefore, we had
to rely on the PQMC results (at $U=4t$) for the gap values
\begin{equation}
\begin{array}{lcl}
\Delta_{1,0}=0.561t, & \quad & \Delta^{0}_{\text{sw}}=1.06t,\\
\Delta^{1}_{\text{sw}}=1.39t, & \quad
&\Delta^{-1}_{\text{sw}}=1.54t,
\end{array}
\label{gapsc60}
\end{equation}
and on the analysis of the molecular orbital energy
level diagram (Fig. 3 of Ref. \onlinecite{lin04}) for the
degeneracies and their ratios
\begin{equation}
\begin{array}{lll}
d_0=1, & d_{1,0}=6, & d_{-1,0}=10, \\
f_{1,0}=120, & f_{-1,0}=10, & f_{0,0}=30.
\end{array}
\end{equation}
%\begin{widetext}
%\begin{eqnarray}
%\ln[Z_{\text{C}}(1)/Z_{\text{C}}(0)]&=&\ln(6e^{-\beta\Delta_{1,0}}+120e^{-\beta\Delta_{\text{sw}}^1})\nonumber\\
%            & &-\ln(1+30e^{-\beta\Delta_{\text{sw}}}),\label{c60lne10}
%            \\
%            \ln[Z_{\text{C}}(-1)/Z_{\text{C}}(0)]&=&\ln(10e^{-\beta\Delta_{-1,0}}+10e^{-\beta\Delta_{\text{sw}}^{-1}})\nonumber\\
%            & &-\ln(1+30e^{-\beta\Delta_{\text{sw}}}),\label{c60lnh10}
%            \\ \nonumber
%\end{eqnarray}
%\end{widetext}
We were able to demonstrate the consistency of the results
obtained by the two methods (ICPQMC and PQMC) by employing a
following procedure. We first let all three gap values be free
fitting parameters. Fitting results are then in agreement with the
PQMC results, albeit with rather large uncertainties. Then we fill
in the two PQMC gap values $\Delta^{\pm 1}_{\text{sw}}$ and
$\Delta^{0}_{\text{sw}}$ [Eq.\ (\ref{gapsc60})], and let
$\Delta_{\pm 1,0}$ be the only free parameter. This procedure, in
general, yields the gap values consistent with the ones previously
obtained using PQMC method \cite{lin04}. Detailed results are
presented in Table \ref{molecules} with representative fits shown
in Fig.\ \ref{c60fitgap}.

\section{Conclusion}
We have generalized the particle-hole symmetric ICPQMC simulation
of Dagotto \emph{et al.}\ \cite{dagotto90} to systems without this
symmetry, such as the tetrahedron, truncated tetrahedron, and
C$_{60}$ molecule. Our simulations show that an accurate canonical
partition function ratio can be obtained through this technique.
Unfortunately, the fitting of these ratios to obtain accurate
energy gaps for C$_{60}$ is impractical. Nevertheless, consistency
between ED, PQMC and ICPQMC has been found.

\begin{acknowledgments}
FL thanks Graeme Luke for helpful advice on figure preparation. We
gratefully acknowledge the support of this project by Natural Sciences
and Engineering Research Council (Canada), The Canadian Institute for
Advanced Research (CIAR), CFI and SHARCNET. All the calculations were
carried out at SHARCNET supercomputing facilities at McMaster
University.
\end{acknowledgments}

\end{document}